\documentclass[12pt]{article}
\usepackage{amssymb}
\usepackage{amsmath}
\usepackage{graphicx}
\usepackage{subfigure}
\usepackage{hyperref}

\begin{document}

\author{Ernst Trojan and George V. Vlasov}
\title{Reply to Comment on ``Acoustics of tachyon Fermi gas''}
\maketitle

\begin{abstract}
In a paper appearing in this issue of Physical Review D, Burmistrov raises
some critical comments on the thermodynamics of a cold tachyon Fermi gas [E.
Trojan and G. V. Vlasov, Phys. Rev. D 83, 124013 (2011)]. However, apart
from any possible theoretical speculation, there are the basic physical
principles 
to test the theory.
\end{abstract}

When one considers a system of many fermions with energy spectrum $%
\varepsilon _k=\sqrt{k^2-m^2}$ and confines himself with the stationary
plane-wave states, corresponding to real energies ($k\geq m$), the 
thermodynamical functions should be operated very carefully.

It is clear that the tachyonic group velocity 
\[
\left| \vec v\right| =\frac{d\varepsilon _k}{dk}=\frac k{\sqrt{k^2-m^2}} 
\]
yields the mean velocity in a many-particle ensemble 
\[
\bar v=\frac{\int\limits_m^{k_F}\left| \vec v\right| \,k^2dk}{%
\int\limits_m^{k_F}k^2dk}=\sqrt{\frac{k_F+m}{k_F-m}}\frac{k_F^2+2m^2}{%
k_F^2+k_Fm+m^2} 
\]
It tends to infinity as soon as the Fermi momentum approaches the lower
cutoff value $k_F\rightarrow m$, implying instability or other anomalous
behavior of the system, particularly, it must concern the sound speed.

The sound speed, determined in our paper \cite{TV2011c} 
\[
c_s=\frac 1{\sqrt{3}}\frac{k_F}{\sqrt{k_F^2-m^2}}
\]
increases unboundedly when $k_F\rightarrow m$. The sound speed suggested in
Comment \cite{B2011} 
\[
c_s=\frac 1{\sqrt{3}}\sqrt{\frac{k_F^2+mk_F+m^2}{k_F^2+mk_F}}
\]
remains always finite, namely, $c_s=1/\sqrt{2}$ at $k_F=m$ that cannot be
realistic because all tachyons of the thermodynamical ensemble are traveling
at infinite high mean velocity. It is clear that our formula reflects the
proper physical behavior of the medium.

We recognize that a contradiction to the third law of thermodynamics $%
E+P=ndE/dn$ is found in our paper \cite{TV2011c}. 
However, it is resolved in an
easier way rather than that proposed in Comment \cite{B2011}. In fact, this
contradiction is no more than a consequence of misprint in unfortunate
Formula (13) in \cite{TV2011c} 
\[
k_F=\left( \frac{6\pi ^2n}\gamma +m^3 \right) ^{1/3}  
\]
The right definition of the Fermi momentum is compatible with the third
law of thermodynamics and it is the following
\begin{equation}
k_F=\left( \frac{6\pi ^2n}\gamma \right) ^{1/3}  \tag{13}
\end{equation} 
where the particle number density equals or exceeds the critical value 
\[
n \geq \frac{\gamma m^3} {6\pi ^2} 
\]
below whom the system is unstable. 
This is the only necessary correction to Ref. \cite{TV2011c} 
but it does not changes the main results, plots and conclusions presented in the article.

{UNPUBLISHED SUPPLEMENTARY NOTE A.}
The tachyonic Dirac sea is depcited in Figure (c), in contrast to the Dirac sea of massive particles (a) and massless particles (b).  

{UNPUBLISHED SUPPLEMENTARY NOTE B.} 
The validity of our thermodynamical approach \cite {TV2011c} with correct formula (13) 
is also evident in the the frames of quantum field theory, 
see the analysis of interacting tachyon Fermi gas [ E. Trojan, arXiv:1204.1370, arXiv:1203.5241].

\begin{figure}[tbp]
\label{dirac1}{\includegraphics[scale=0.35]{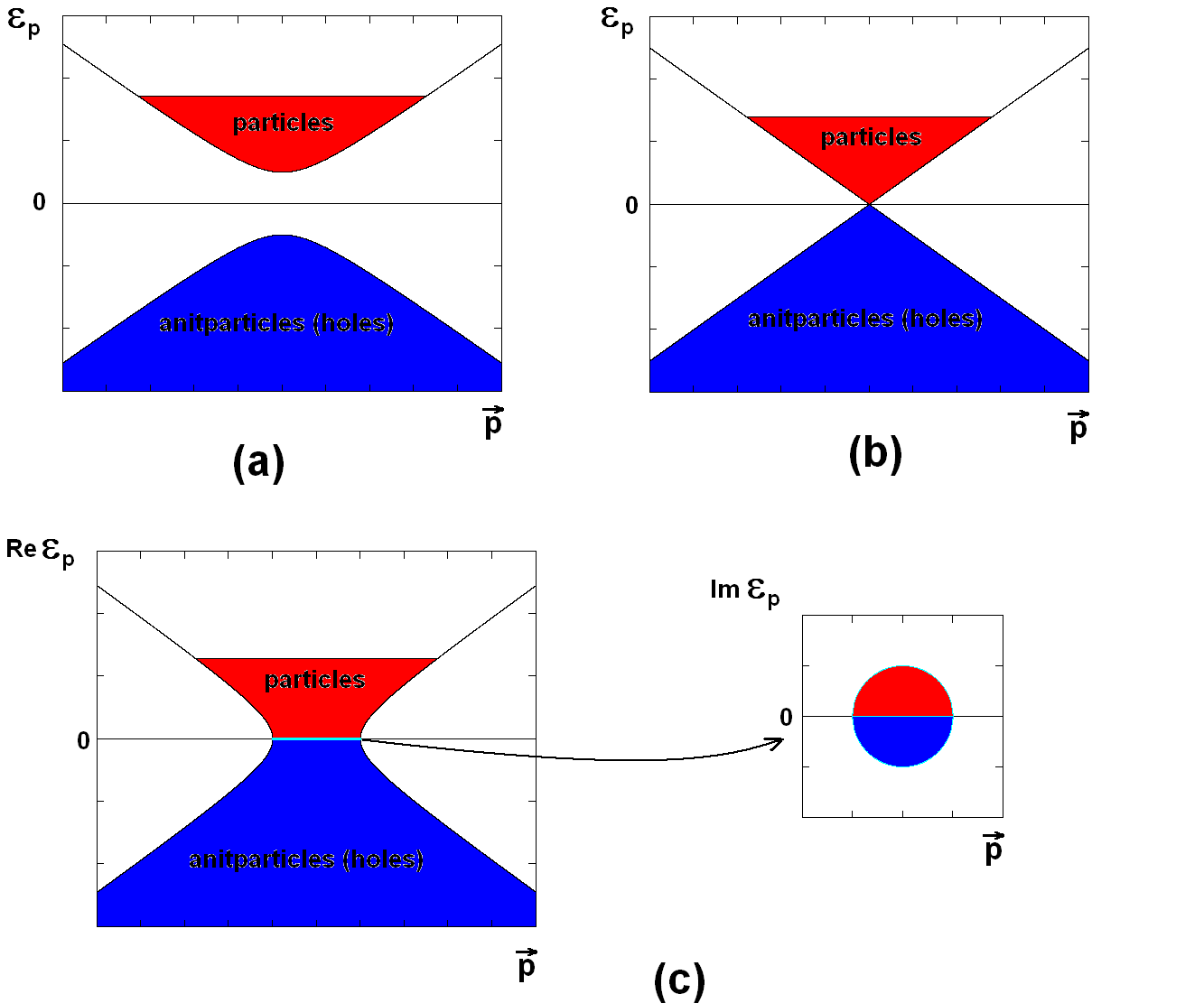}}
\end{figure}

\end{document}